\documentclass{WileyChemistry-template}
\usepackage{amsmath}
\usepackage{mathtools,amssymb,lipsum, nccmath}

\usepackage{cuted}
\title{Probe Beam Dichroism and Birefringence in Stimulated Raman Scattering in Polyatomic Molecules}

\author{
\begin{minipage}{\textwidth}
	Bogdan V. Semak,\textsuperscript{+,[a]} Yaroslav M. Beltukov,\textsuperscript{+,[b]} Oleg S. Vasyutinskii*\textsuperscript{+,[c]}
\end{minipage}
}

\newcommand{\affiliation}{
\begin{itemize}

\item[{[a]}] Bogdan V. Semak \\
Ioffe Institute, Russian Academy of Sciences, Polytekhnicheskaya 26, 194021 St.Petersburg, Russia

\item[{[b]}] Dr. Yaroslav M. Beltukov\\
{Peter the Great St. Petersburg Polytechnic University, Polytechnicheskaya 29,
195251 St. Petersburg;\\
Ioffe Institute, Russian Academy of Sciences, Polytekhnicheskaya 26, 194021 St.Petersburg, Russia
}

\item[ {[c]}] Prof. Oleg S. Vasyutinskii*\\
osv@pms.ioffe.ru \\
Ioffe Institute, Russian Academy of Sciences, Polytekhnicheskaya 26, 194021 St.Petersburg, Russia

\item[{[\texttt{+}]}] These authors contributed equally.
\end{itemize}
}

\renewcommand{\dedication}{
	\begin{minipage}{\textwidth}
		Dedication (Special Collection: Pablo Villarreal Herrán Festschrift)
	\end{minipage}
}

\renewcommand{\abstract}{Dichroism and birefringence in Stimulated Raman Scattering (SRS) in polyatomic molecules were studied theoretically. General expressions describing the change of the polarization matrix of the probe laser beam transmitted through initially isotropic molecular sample excited by the pump laser beam have been derived. Arbitrary  polarization states and propagation directions of the incoming pump and probe beams were considered. The expressions were written in terms of spherical tensor operators that allowed for separation of the field polarization tensor and the molecular part containing three scalar values of nonlinear optical susceptibility $\chi^{(3)}_{K_{pu}}$ with $K_{pu}$  = 0,1,2. The geometry of almost collinear propagation of the pump and probe beams through the molecular sample was considered in greater details. It was shown that the dichroism and birefringence refer to the nonlinear optical susceptibility element $\chi^{(3)}_{2}$  and that their contributions to the SRS signal  can be separated experimentally by using an appropriate probe beam polarization analyzer installed in front of the photodetector. Particular cases of the off-resonant SRS and resonant SRS have been considered. The results obtained were expressed in terms of the Stokes polarization parameters of the pump and probe beams. }

\newcommand{\keywords}{
	Keyword 1 \textbullet\
	Keyword 2 \textbullet\
	Keyword 3 \textbullet\
	Keyword 4 \textbullet\
	Keyword 5
}

\begin{document}

\twocolumn[\vspace{-1.5cm}\maketitle\vspace{-1cm}
	\textit{\dedication}\vspace{0.4cm}]
\small{\begin{shaded}
		\noindent\abstract
	\end{shaded}
}

\begin{figure} [!b]
\begin{minipage}[t]{\columnwidth}{\rule{\columnwidth}{1pt}\footnotesize{\textsf{\affiliation}}}\end{minipage}
\end{figure}




\section*{Introduction}
\label{introduction}

Stimulated Raman Scattering (SRS) spectroscopy in particular SRS microscopy has nowadays found  wide applications for analysis and label-free imaging of functional endogenous biomolecules in living systems as it offers critical information to understand the fundamentals in biology and to assist clinical diagnostics~\cite{Prince2017,Shipp2017,Zhang2018,Zhang2021}. An advantage of SRS microscopy in comparison with spontaneous Raman microscopy is the dramatic increase of experimental signals and improvement in imaging speed, which gives rise to real-time hyperspectral vibrational imaging of live biological samples. Another popular coherent Raman scattering technique is anti-Stokes Raman scattering (CARS) that is also widely used in life science applications~\cite{Shipp2017} is usually accompanied by a nonresonant background, resulting from nonlinear optical responses mediated through molecular virtual or electronic states, or both~\cite{Cheng2003} while in SRS the nonresonant background is usually sufficiently reduced allowing for rapid hyperspectral imaging at low analyte concentrations~\cite{Stolow2015}.

Both techniques are based on the interaction of laser beams at pump frequency $\omega_p$ and Stokes frequency $\omega_s$ when the frequency difference $\Delta\omega = \omega_p-\omega_s$ is equal to the resonant frequency of a vibrational transition within the molecular sample. The SRS signal is generated at the frequency of an incident beam and is based on the energy transfer between the pump and the Stokes beams where the measured signal may either be a power gain in the Stokes beam (Stimulated Raman Gain (SRG)) or a loss in the pump beam (Stimulated Raman Loss (SRL)).

As known~\cite{Mukamel95,Boyd20}, all information on the dynamics of third-order nonlinear optical phenomena is contained in the fourth-order tensor of nonlinear optical susceptibility $\chi^{(3)}_{ijkl}(\omega_m,\omega_n,\omega_q)$, where the subscript indices $i,j,k,l$ are Cartesian directions of light polarizations and $\omega_m,\omega_n,\omega_p$ are light frequencies involved in the nonlinear scattering process. The nonlinear susceptibility performs the relationship between the components  of nonlinear polarization $P_i(\omega)$ generated in the sample and the Fourier components of the light electric vectors~\cite{Boyd20}, where $\omega$ is the frequency of the outgoing detected light beam.

The third-order nonlinear susceptibility $\chi^{(3)}_{ijkl}(\omega_m,\omega_n,\omega_q)$ contains the matrix elements of molecular transition dipole moments and the terms that include transition frequencies and the frequencies of light electric fields. The number of independent matrix elements depends on the sample symmetry, in particular in isotropic media in solutions and in certain biological samples the number of third-order nonlinear susceptibility independent matrix elements is limited to three~\cite{Mukamel95,Boyd20}.

Although the general quantum mechanical theory of SRS and other third-order nonlinear optical phenomena in any media is well developed~\cite{Mukamel95,Boyd20} and particular experimental geometries were considered~\cite{Xie2011,Stolow2015,Wang2016} the detailed theory of possible polarization effects under SRS in biologically relevant polyatomic molecules was not exist till now.

This paper aims to address this problem. General expression for the change of the polarization matrix of the probe (Stokes) laser beam transmitted through the ensemble of polyatomic molecules excited by a pump laser beam. {The molecular sample was assumed to be initially isotropic, that is its dichroism and birefringence was entirely due to interaction with two laser beams.} The expression is presented also in terms of Stockes parameters of the incoming and outgoing pump and probe laser beams. The contributions of linear dichroism and birefringence to the change in the probe beam polarization matrix were discussed. {The geometry of almost collinear propagation of the pump and probe beams through the molecular sample was considered in greater details.}

The spherical tensor approach was used throughout the paper for description of the molecular and light polarization matrices~\cite{Zare88b,Blum96}. Similar approach was used to describe CARS before~\cite{Cleff2017} where however no specification for excitation of polyatomic molecules and no general expression and its analysis were presented.  The paper extends the approach used in our recent publications~\cite{Shternin10,Denicke10,Gorbunova20b,Semak21} to describe the SRS of polyatomic asymmetric top molecules in the condensed phase.

The main results obtained are as follows. As shown,  {the outgoing beam polarization matrix in the spherical tensor representation (polarization tensor) can be written as a product of the polarization-dependent field tensor and the polarization-independent molecular part containing three scalar third-order nonlinear susceptibility terms $\chi^{(3)}_0$, $\chi^{(3)}_1$, and $\chi^{(3)}_2$ having each clear physical meanings. The field tensor can be completely controlled by experimentalist that allows for analysis of an arbitrary experimental SRS geometry including all possible light beams propagation directions and polarizations.}

{The components of the polarization tensor of an elliptically polarized light beam propagating along a direction $\mathbf{n}$  have been tabulated. Particular expressions for  absorption/gain intensity, rotation of the polarization plane, and  birefringence of the probe beam  in terms of the third-order nonlinear susceptibilities $\chi^{(3)}_0$, $\chi^{(3)}_1$, and $\chi^{(3)}_2$ have been written. The features of the general expression for the third-order nonlinear susceptibility $\chi^{(3)}_K$, where $K=0,1,2$  have been analyzed. Particular cases of off-resonance and resonance SRS have been considered.}



\section*{Results and Discussion}
\label{results_discussion}

\subsection*{Description of the Detected Light Polarization}
\label{Description of light}

We consider the experimental scheme presented in Scheme~\ref{fig1} where the pump and probe light beams $I_{pu}$ and $I_{pr}$ propagate along axis Z almost collinearly through the molecular sample (MS). The angle between the beams in Scheme~\ref{fig1} is greatly increased for the sake of clarity. It is assumed that only the probe light beam is recorded by the photodetector (D), while the pump beam passed through MS is somehow blocked.  In the case of SRS the probe Stokes beam frequency $\omega_{pr}=\omega_S$ is smaller than the pump beam frequency $\omega_{pu}$~\cite{Boyd20}. However for generality we will also consider the case when the probe anti-Stokes $\omega_{pr}=\omega_{AS}$ frequency is larger than the pump frequency $\omega_{pu}$. The beams $I_{pu}$ and $I_{pr}$  are assumed to be  pulsed with a time delay $\tau$ between them. Initial beam polarizations are controlled by polarizers P$_1$ and P$_2$ and the output polarization of the probe beam is analysed by a polarization analyzer A and recorded by a photodetector. The polarization of each beam can be chosen arbitrarily (linear, circular, elliptical).

\begin{scheme}
    \centering
    \includegraphics[width=8.6cm]{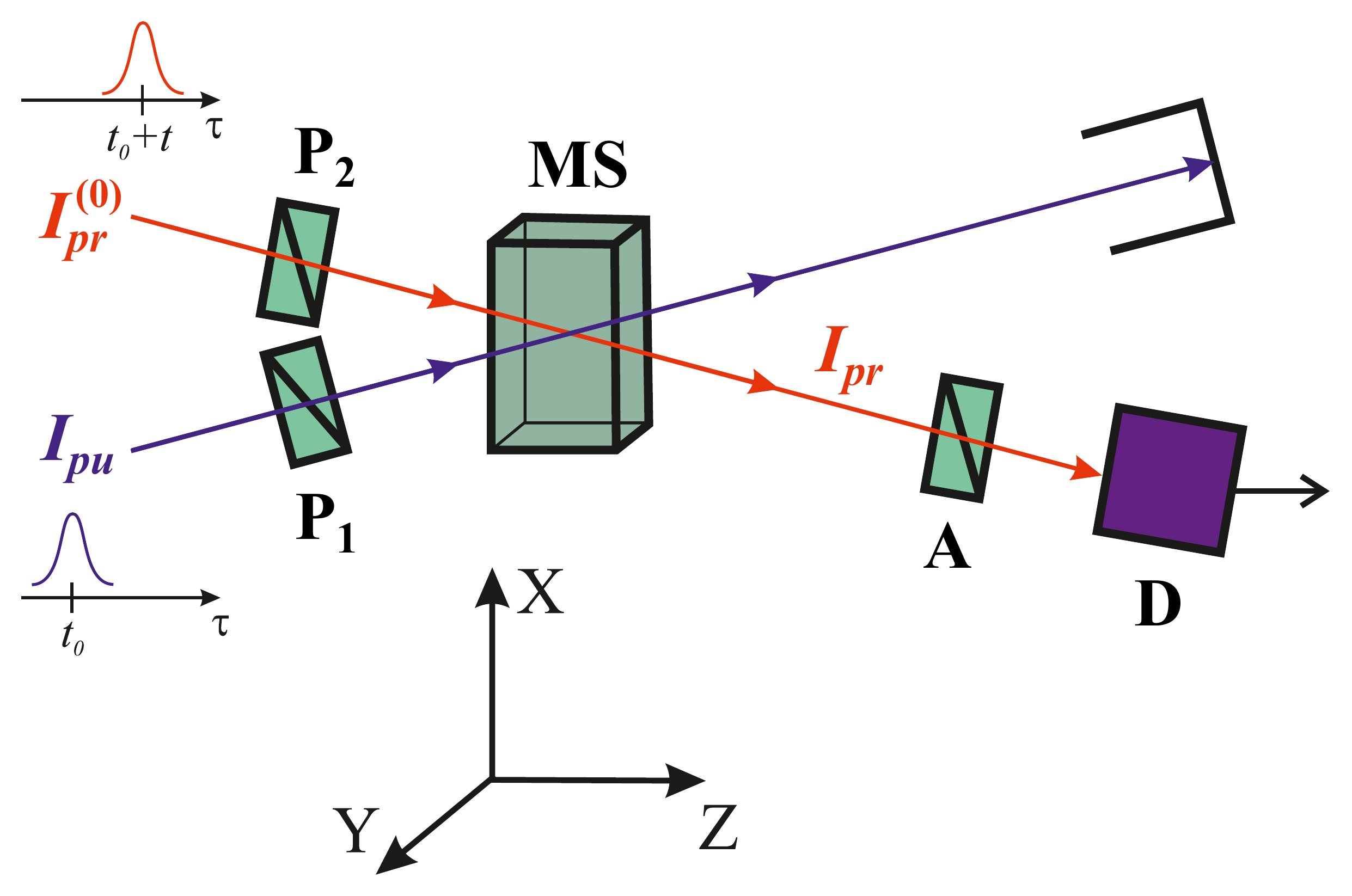}
    \caption{Generalized SRS scheme. \\
    $\mathrm{MS}$ is a molecular sample, $I_{pu}$ and $I_{pr}$ are the pump and probe beams, respectively, $\mathrm{P}_1$ and $\mathrm{P}_2$ are polarizers, $\mathrm{A}$ is a polarization analyser, $\mathrm{D}$ is a photodetector.}
\label{fig1}
\end{scheme}

The light polarization matrix of each of the beams in Scheme~\ref{fig1} are given by
\begin{equation}
\label{eq:pol_matrixii'}
\pi_{ii'}(\omega) = E_i(\omega) E^{*}_{i'}(\omega),
\end{equation}
where $E_i(\omega)$ is the amplitude of a Cartesian electric field component at the frequency $\omega$ and $i,i'=x,y$.

The light beam intensity $I(\omega)$ is given by:
\begin{equation}
\label{eq:Intensity}
I(\omega)=\frac{C}{2}\mathrm{Tr}[\pi(\omega)],
\end{equation}
where $C= c\epsilon_0 n $, $\epsilon_0$ is the electric constant, $n$ is the refraction index at the frequency $\omega$, and $c$ is the speed of light in vacuum.

The light polarization matrix $\pi$  in eq.~(\ref{eq:pol_matrixii'}) can be rewritten in terms of the Stokes polarization parameters~\cite{Collett2005,Chen2010} $S_1$, $S_2$, and $S_3$ :
\begin{align}
\label{Stokes}
\pi(\omega)=\frac{I}{C}\left( \begin{array}{ccc} 1+S_1  & S_2-iS_3 \\ S_2+iS_3 & 1-S_1 \end{array} \right),
\end{align}
where $i$ is the imaginary unit.

The particular cases (a) $I_1=\pm1, I_2=0, I_3=0$, (b) $I_1=0, I_2=\pm1, I_3=0$, and (c) $I_1=0, I_2=0, I_3=\pm1$ relate to: (a) light is linearly polarized along X/Y axis; (b)  light is linearly polarized on 45$^{\circ}$/135$^{\circ}$ to axis X; (c) light is right/left handed circularly polarized. The Stokes polarization parameters are very important because they can easily be measured experimentally.

General expression for the change of the probe light polarization matrix  $\pi$ within SRS was obtained in the frame of the third-order perturbation theory. The Fourier transform of the third-order nonlinear polarization component $P^{(3)}_i(\omega)$  at the frequency $\omega$ can be presented in the form~\cite{Mukamel95,Boyd20}:
\begin{multline}
\label{eq:3rdOrdFourier}
P^{(3)}_k(\omega)=
\epsilon_0
\sum_{j,i,h}
\int\limits^{+\infty}_{-\infty} d\omega_3\int\limits^{+\infty}_{-\infty} d\omega_2\int\limits^{+\infty}_{-\infty} d\omega_1 \\
\times
E_j(\omega_1)E_i(\omega_2)
E_h(\omega_3)\, \\
\times
\chi^{(3)}_{kjih}(\omega; \omega_1,\omega_2,\omega_3) \,\delta(\omega_1+\omega_2+\omega_3-\omega),
\end{multline}
where the indices $k,j,i,h$  refer to the Cartesian components of the fields, $\chi^{(3)}_{kjih}(\omega; \omega_1,\omega_2,\omega_3)$ is the third-order nonlinear susceptibility, and $\delta(\omega_1+\omega_2+\omega_3-\omega)$ is the Delta-function that leads to the frequency rule: $\omega_1+\omega_2+\omega_3=\omega$.

The nonlinear susceptibility  obeys the intrinsic permutation symmetry meaning that it  is unchanged by the simultaneous interchange of the frequencies $\omega_1,\omega_2,\omega_3$ and the corresponding Cartesian indices~\cite{Boyd20}. It also obeys a symmetry rule~\cite{Boyd20}:
\begin{equation}
\label{eq:symmetry}
\chi^{(3)*}_{kjih}(\omega; \omega_1,\omega_2,\omega_3)=\chi^{(3)}_{kjih}(-\omega; -\omega_1,-\omega_2,-\omega_3).
\end{equation}
The Fourier  components of electric field $E_j(\omega_1)$, $E_i(\omega_2)$, and $E_h(\omega_3)$ in eq.~(\ref{eq:3rdOrdFourier}) are in general complex values that
%
%
obey the usual symmetry rule~\cite{Boyd20}:
 \begin{equation}
\label{eq:Ei}
E^*_j(\omega) = E_j(-\omega).
\end{equation}

For each beam, the Fourier components can be written in the form:
\begin{align}
\label{eq:env_pu}
E_j(\omega_{pu}) &= {\cal E}(\omega_{pu})\,\mathbf{e}_j^{pu}, \\
\label{eq:env_pr}
E_j(\omega_{pr}) &= {\cal E}(\omega_{pr})\,\mathbf{e}_j^{pr},
\end{align}
where ${\cal E}(\omega_{pu})$ and ${\cal E}(\omega_{pr})$ describe the spectrum of the corresponding beam. 
In the following we will assume that the pump and the probe beams are characterized by relatively narrow frequency envelopes with the widths $\Delta_{pu}$, $\Delta_{pr}$ around their carrier frequencies $\omega^{(c)}_{pu}$ and $\omega^{(c)}_{pr}$ and that within each beam the light polarization vector $\mathbf{e}_j$ in eqs.~(\ref{eq:env_pu}), (\ref{eq:env_pr}) does not depend on the frequency $\omega$.



In general, all frequencies in eq.~(\ref{eq:3rdOrdFourier}) may be pump or probe frequencies. In the case of SRS that is a subject of this paper, two of $\omega_1,\omega_2,\omega_3$ are pump frequencies. The remaining frequency within $\omega_1,\omega_2,\omega_3$ and the frequency $\omega$ are probe frequencies. Thus, we make the association $\omega=\omega_{pr}$, $\omega_1=\omega'_{pr}$, $\omega_2=-\omega'_{pu}$, and $\omega_3=\omega_{pu}$. Since there are six equal possibilities to make such association, we add the factor 6 in eq.~(\ref{eq:A(Kpu)}) below.

According to the theory the Fourier transform of a Cartesian component of the outgoing probe electric field $E^{out}_k(\omega_{pr})$  can be presented as a sum~\cite{Domcke1992,Mukamel95}:
\begin{equation}
\label{eq:Eout}
    E^{out}_k(\omega_{pr}) = E^{in}_k(\omega_{pr}) + \frac{i\,\omega^{(c)}_{pr}\ell}{2n\epsilon_0c}P^{(3)}_k(\omega_{pr}),
\end{equation}
where $E^{in}_k(\omega_{pr})$  is an electric field  component in the incoming probe beam at frequency $\omega_{pr}$, $\omega^{(c)}_{pr}$ is the probe beam carrier frequency, $\ell$ is the MS length,  and $P^{(3)}_k(\omega_{pr})$ is the Fourier transform of a third-order nonlinear polarization component.

Equation (\ref{eq:Eout}) is valid when the probe beam intensity is much weaker than the pump beam intensity and in the slowly varying amplitude approximation~\cite{Mukamel95} meaning that the probe field envelope varies slowly in time compared with its optical period. Also, thin optical layer condition $\alpha \ell \ll 1$  was assumed, where $\alpha$ is  absorption/gain coefficient of the probe beam.

The spherical tensor representation of the light polarization matrix is used throughout this paper for all electric fields involved in eqs.~(\ref{eq:Eout}) and (\ref{eq:3rdOrdFourier}). The relationship between the Cartesian $\mathbf{e}_i$ and spherical $\mathbf{e}_p$ unit vectors is given by~\cite{Varsh88}:
\begin{align}
\label{eq:spherical}
\mathbf{e}_1&=-\frac{1}{\sqrt2}(\mathbf{e}_x+i\mathbf{e}_y) \nonumber  \\
\mathbf{e}_{-1}&=\frac{1}{\sqrt2}(\mathbf{e}_x-i\mathbf{e}_y) \nonumber \\
\mathbf{e}_{0} &= \mathbf{e}_z.
\end{align}

{Using eq.~(\ref{eq:spherical}) the covariant components of the light polarization matrix in the spherical tensor representation that is known as polarization tensor can be defined as~\cite{Happer72,Zare88b,Picheyev97}:}
\begin{equation}
\label{eq:EKQ}
    E_{KQ}(\mathbf{e}) = \sum\limits_{p, p'} (-1)^{p'} C^{K \, Q}_{1 \, p\: 1 \, -p'} \mathbf{e}_{p} \mathbf{e}^*_{p'} ,
\end{equation}
where $p,p'=0,\pm1$ are spherical indices and the term $C^{K \, Q}_{1 \, -p\: 1 \, p'}$ is a Clebsch-Gordan coefficient~\cite{Varsh88}.

The indices $K$ and $Q$ in eq.~(\ref{eq:EKQ}) are the light polarization rank and its component onto the laboratory axis Z. The rank $K$ can take the values $K=0,1,2$.

The  tensor component $E_{00}=-1/\sqrt3$ is proportional to an isotropic part of light (normalised total light intensity) while the tensor components $E_{1Q}$ with $Q=0,\pm1$ and $E_{2Q}$ with $Q=0,\pm1,\pm2$ describe various types of light polarization. In particular, the tensor components $E_{1Q}$ describe the \emph{light helicity}  (light spin orientation) and refer to light circular polarization while the tensor components $E_{2Q}$ describe the light electric vector alignment and refer to light linear polarization.

The $E_{KQ}(\mathbf{e})$ tensor components in eq.~(\ref{eq:EKQ}) can be expressed in terms of the Stokes parameters as follows~\cite{Picheyev97}:
\begin{align}
\label{eq:Stokes2}
E_{00}(\mathbf{e}) =-\frac{1}{\sqrt3}; \: E_{10}(\mathbf{e}) =\frac{S_3}{\sqrt2}; \: E_{20}(\mathbf{e}) =-\frac{1}{\sqrt6}  \nonumber \\
E_{22}(\mathbf{e})+E_{2-2}(\mathbf{e})=S_1; \: E_{22}(\mathbf{e})-E_{2-2}(\mathbf{e})=iS_2.
\end{align}


\subsection*{Description of the Molecular Sample}
We assumed that the molecular system interacting with light was coupled to a thermal bath and could be described by a reduced equation of motion for the density operator. A simple form of reduced equations of motion was adopted~\cite{Mukamel95}:
\begin{align}
\label{eq:equation motion}
\frac{d}{dt}\rho_{\nu,\nu'}=(-i\omega_{\nu,\nu'}-\Gamma_{\nu,\nu'})\rho_{\nu,\nu'}
\end{align}
with
\begin{align}
\label{eq:relax}
\Gamma_{\nu,\nu'} = \frac{1}{2}(\gamma_{\nu}+\gamma_{\nu'})+\hat{\Gamma}_{\nu,\nu'},
\end{align}
where $\rho_{\nu,\nu'}$ is the matrix elements of the reduced density operator traced over the bath, $\gamma_{\nu}$ and $\gamma_{\nu}$ are the inverse lifetimes of the $\nu$ and $\nu'$ energy states, and $\hat{\Gamma}_{\nu,\nu'}$ is the pure dephasing rate for the $\nu,\nu'$ transitions.

We considered the ensemble of polyatomic molecules  that has a multilevel Hamiltonian $H_{\nu,\nu'}$  that includes electronic, vibrational, and rotational degrees of freedom. The total molecular wave function in the Born-Oppenheimer approximation was written in the form:
\begin{align}
\label{E:Psimol}
\Psi^{tot}=\Psi_n(\mathbf{r}_e)^{el}\Psi_{v}(R)\Psi^{rot}_{J,M,\tau}(\alpha,\beta,\gamma),
\end{align}
where $n$ is the aggregate of electronic quantum numbers, $\mathbf{r}_e$ is the aggregate of all electron coordinates, $v$ are vibrational quantum numbers, $R$ is the aggregate of all nuclear coordinates, $J$  and $M$  are the total molecular angular momentum and its projection onto the laboratory axis Z, $\tau$ are rotational asymmetric top quantum numbers~\cite{Zare88b}, and $(\alpha,\beta,\gamma)$ are Euler angles describing molecular position in space.

The rotational wave function was presented as an expansion over symmetric top wave functions~\cite{Zare88b}:
\begin{align}
\label{E:A}
\Psi^{rot}_{J,M,\tau}(\alpha,\beta,\gamma)  = \sum_{\Omega}A^{J}_{\tau \Omega} D^{J^*}_{M,\Omega}(\alpha,\beta,\gamma),
\end{align}
where $A^{J}_{\tau \Omega}$ are expansion coefficients and $D^{J}_{M,\Omega}(\alpha,\beta,\gamma)$ are Wigner $D$-functions.

\subsection*{General Expression for the Probe Beam Polarization Tensor $E_{KQ}(\mathbf{n})$}

The general expression for the probe beam polarization tensor below was derived assuming initially isotropic molecular sample, and particularly isotropic distribution of the total angular momenta $J_g$ in the ground electronic molecular state. We also assumed that the duration of the pump and probe light pulses is much shorter than any reorientational molecular motions that relates to the condition of excitation with femtosecond laser pulses.

Then, combining eqs.~(\ref{eq:pol_matrixii'})--(\ref{eq:Eout}), (\ref{E:Psimol}), (\ref{E:A}) and using explicit expressions for the third-order nonlinear susceptibility $\chi^{(3)}_{ijkh}(\omega; \omega_{1}, \omega_{2}, \omega_{3})$ from~\cite{Mukamel95,Boyd20} after cumbersome transformations using angular momentum algebra~\cite{Varsh88} and including averaging over molecular axes directions, transformation of transition matrix elements to the molecular frame and summation over all angular momenta and their projections one can obtain the following expression for the change of the probe beam polarization tensor:

\begin{multline}
\label{eq:general}
    I^{out}(\omega_{pr})E^{out}_{KQ}(\mathbf{e}_{pr}) - I^{in}(\omega_{pr})E^{in}_{KQ}(\mathbf{e}_{pr}) \\
     =\sum_{K_{pu},K_{pr}} (2K_{pu}+1)^{1/2}(2K_{pr}+1)^{1/2} \,(-1)^{K_{pu}+ K_{pr}}
         \\
        \times
         \left\{
        \begin{array}{ccc}
            K & K_{pu} & K_{pr} \\
            1 & 1      & 1
        \end{array}
        \right\}
         \left[E_{K_{pu}}(\mathbf{e}_{pu}) \otimes E^{in}_{K_{pr}}(\mathbf{e}_{pr}) \right]_{KQ}  \\
         \times
         \left(A^{}_{K_{pu}}(\omega_{pr}) + (-1)^{K+K_{pu}+ K_{pr}}A^{*}_{K_{pu}}(\omega_{pr})\right),
\end{multline}
where the term in curly brackets is a 6-$j$ symbol and the irreducible tensor product is defined according to~\cite{Varsh88}
\begin{multline}
     \left[E_{K_{pu}}(\mathbf{e}_{pu}) \otimes E^{in}_{K_{pr}}(\mathbf{e}_{pr}) \right]_{KQ}  \\
    = \sum_{Q_{pr},Q_{pu}} C^{K Q}_{K_{pu} Q_{pu} \, K_{pr}  Q_{pr}}
    E_{K_{pu} \, Q_{pu}}(\mathbf{e}_{pu}) E^{in}_{K_{pr} \, Q_{pr}}(\mathbf{e}_{pr}).
\end{multline}

The ranks $K_{pu}$ and $K_{pr}$  in eq.~(\ref{eq:general}) can in general take the values: $K_{pu}, K_{pr}$\,=\,0,1,2 and fully characterise the pump and probe pulse polarizations and the polarization of MS described by the coefficients $A^{}_{K_{pu}}(\omega_{pu})$. As can be seen in eq.~(\ref{eq:general})  the coefficients $A^{}_{K_{pu}}(\omega_{pu})$  depend on the rank $K_{pu}$ of the pump pulse polarization and in general can take only three values $A^{}_{0}(\omega_{pu})$, $A^{}_{1}(\omega_{pu})$ , and $A^{}_{2}(\omega_{pu})$   at each frequency $\omega_{pu}$  describing population, orientation, and alignment of MS, respectively.

The coefficients $A^{}_{K_{pu}}(\omega)$ in eq.~(\ref{eq:general}) are given by:
\begin{equation}
\label{eq:A(Kpu)}
    A_{K_{pu}}(\omega_{pr}) = 3 i\, \omega^{(c)}_{pr}\ell\,{\cal E}^*(\omega_{pr}) \,P^{(3)}_{K_{pu}}(\omega_{pr}),
\end{equation}
where $P^{(3)}_{K_{pu}}(\omega_{pr})$  is the third-order  nonlinear polarization of the MS:
\begin{multline}
\label{eq:P(Kpu)}
    P^{(3)}_{K_{pu}}(\omega_{pr}) = \iiint
    d\omega_{pr}'d\omega_{pu}'d\omega_{pu}   \\
    \times
     {\cal E}(\omega_{pr}')
    {\cal E}^*(\omega_{pu}')
    {\cal E}(\omega_{pu})
    \chi^{(3)}_{K_{pu}}(\omega_{pr};\omega_{pr}',-\omega_{pu}',\omega_{pu})\\
    \times \delta(\omega_{pr}-\omega_{pr}'+\omega_{pu}'-\omega_{pu}),
\end{multline}
where $ \chi^{(3)}_{K_{pu}}(\omega_{pr};\omega_{pr}',-\omega_{pu}',\omega_{pu})$ is the third-order nonlinear susceptibility presented in the spherical tensor form.

As can be seen in eq.~(\ref{eq:P(Kpu)}) the nonlinear polarization and nonlinear susceptibility in their spherical tensor form depend on the rank $K_{pu}$ of the pump light and on the frequencies  of the applied fields, but not on any projections. The zero-rank MS polarization $P^{(3)}_{0}$ refers to the isotropic distribution of molecular axes and is responsible to absorption/gain of the probe beam that does not depend on the pump beam polarization. The first-rank MS polarization $P^{(3)}_{1}$ refers to the molecular axes orientation, and the second-rank MS polarization $P^{(3)}_{2}$ refers to the molecular axes alignment.

The coefficients $A^{}_{K_{pu}}(\omega_{pr})$ in eqs.~(\ref{eq:general}) and (\ref{eq:A(Kpu)}) are in general complex functions (in the mathematical sense). The phase $(-1)^{\phi}=(-1)^{K+K_{pu}+K_{pr}}$ in eq.~(\ref{eq:general}) is very important as it describes various types of signals that can be observed. If the phase $(-1)^{\phi}$ is positive then the last line in the right hand side (\emph{rhs}) of eq.~(\ref{eq:general}) is real, while if it is negative then the last line in the  \emph{rhs} is pure imaginary. As will be shown below the real value of the \emph{rhs} in eq.~(\ref{eq:general}) describes the probe beam loss (or gain) and dichroism  while the imaginary part  of the \emph{rhs} describes the probe beam birefringence. Therefore, experimentalist can control these two regimes by changing the incoming, or outgoing light beams polarization.

The expression for the change of the probe beam polarization tensor $E^{}_{KQ}(\omega_{pr})$ in eq.~(\ref{eq:general}) has undeniable advantages as it allows to describe arbitrary elliptical polarizations of the probe and pump beams and in general any relative directions of the beams. Also, the first and the second lines in  eq.~(\ref{eq:general}) describe light polarization contribution to the signal that can be controlled by experimentalist while the third line describes the contribution from the material part that describes the properties of the MS. As will be shown later this separation greatly simplifies the signal analysis.  Moreover, the spherical tensor approach allows to obtain a close expression for the third-order nonlinear susceptibility $\chi_{K_{pu}}(\omega_{pr};\omega_{pr}',-\omega_{pu}',\omega_{pu})$ in eq.~(\ref{eq:A(Kpu)}) for arbitrary polyatomic molecules with any angular momentum values.

\subsection*{Third-Order Nonlinear Susceptibility}
\label{sec:susceptibility}

The general expression for the nonlinear susceptibility in the spherical tensor representation $\chi^{(3)}_{K_{pu}}(\omega_{pr};\omega_{pr}',-\omega_{pu}',\omega_{pu})$  in eq.~(\ref{eq:P(Kpu)}) is given by:
\begin{multline}
\label{eq:chi(3)}
    \chi^{(3)}_{K_{pu}}(\omega_{pr};\omega_{pr}',-\omega_{pu}',\omega_{pu}) \\
    = \frac{N}{ \hbar^3}
   \sum_{q_1q_2}\sum_{q_3q_4}\,\frac{(-1)^{q}}{2K_{pu}+1}\,  C_{1\, q_3\,1\,q_4}^{K_{pu}q}\,C_{1\,-q_2\,1\,-q_1}^{K_{pu}q}\\
    \times
    \left( R_{\tilde q}(\tilde\omega)+ \tilde R_{ q}^*(-\tilde\omega)\right),
\end{multline}
where $N$ is the molecular concentration.

The indices $q_1,q_2,q_3,q_4$  in eq.~(\ref{eq:chi(3)}) label the spherical components of molecular transition dipole moments in the molecular frame, $(\tilde\omega)$ is the aggregate of the applied fields frequencies $(\omega_{pr}, \omega'_{pr}, -\omega'_{pu}, \omega_{pu})$, and $\tilde q$ is the aggregate of the indices $q_1,q_2,q_3,q_4$.
The function $ \tilde R_{\tilde q}(\tilde\omega)$ is given by:

\begin{multline}
\label{eq:R}
 R_{\tilde q}(\tilde\omega)={\cal \hat P}_I\sum_{gee'f} \rho_{gg}^{(0)} I_{gee'f}(\omega_{pr}; \omega'_{pr}, -\omega_{pu}', \omega_{pu})
 \\
      \times
  D_{\tilde q}(g,e,e',f),
\end{multline}
where $\rho^{(0)}_{gg}$ denotes the density matrix of vibrational energy states in the molecular ground state and the function $D_{\tilde q}(g,e,e',f)$ is
\begin{equation}
\label{eq:D}
 D_{\tilde q}(g,e,e',f)=
      \langle g | d_{q_1} | e \rangle \langle e | d_{q_2} | f \rangle
    \langle f | d_{q_3} | e' \rangle \langle e' | d_{q_4} | g \rangle,
\end{equation}
where $\langle g | d_{q} | e \rangle=d^{q}_{ge}$ are transition matrix elements written in molecular frame.

The intrinsic permutation operator ${\cal \hat P}_I$ in eq.~(\ref{eq:R}) tells to average the expression that follows it over all permutations
of the frequencies $\omega_{pu}$, $-\omega'_{pu}$, and $\omega'_{pr}$. The spherical indices $q_i$ in the transition dipole matrix elements, where $i=1,2,3,4$ are to be permuted simultaneously. The wave functions $|g\rangle$, $|e\rangle$, $|e'\rangle$, and $|f\rangle$ contain the electronic and vibrational quantum numbers, but not any rotational quantum numbers. Note that the function $ R_{\tilde q}(\tilde\omega)$ in eq.~(\ref{eq:chi(3)}) written in the spherical tensor basis is similar to the function $R^{ijkl}_{\alpha}$ introduced by Mukamel~\cite{Mukamel95} in the Cartesian basis however not to be confused with it. The function $ I_{gee'f}(\omega_{pr}; \omega_{pr}',-\omega_{pu}',\omega_{pu})$ in eq.~(\ref{eq:R}) is given by:

\begin{align}
\label{eq:I_gee'f}
    &I_{gee'f}(\omega_{pr}; \omega_{pr}',-\omega_{pu}',\omega_{pu})  \nonumber \\
    &=\Big(I_{eg}(\omega_{pr}) I_{fg}(\omega_{pr}'-\omega_{pu}') I_{e'g}(\omega_{pr'})\nonumber\\
     &+I_{e'f}(\omega_{pr})  I_{gf}(\omega_{pr}'-\omega_{pu}') I_{ge}(-\omega_{pu}')\nonumber \\
     &+I_{e'f}(\omega_{pr})  I_{e'e}(\omega_{pu}-\omega_{pu}') I_{e'g}(\omega_{pu})\nonumber \\
    &+I_{e'f}(\omega_{pr})  I_{e'e}(\omega_{pu}-\omega_{pu}') I_{ge}(-\omega_{pu}')
   \Big).
\end{align}
where the functions $I_{ab}(\omega)$ are defined as~\cite{Mukamel95}:
\begin{equation}
\label{eq:Iab(omega)}
    I_{ab}(\omega) = \frac{1}{\omega - (\omega_a - \omega_b) + i\Gamma_{ab}}.
\end{equation}
The function $I_{ab}(\omega)$ in eq.~(\ref{eq:Iab(omega)}) obeys the symmetry rule:
\begin{equation}
\label{eq:Iab(omega)sym}
    I^*_{ab}(-\omega) = - I_{ba}(\omega).
    \end{equation}



{The irreducible form of the nonlinear susceptibility $\chi^{(3)}_{K_{pu}}(\omega_{pr};\omega_{pr}',-\omega_{pu}',\omega_{pu})$ in eq.~(\ref{eq:chi(3)}) depends on the symmetry of molecular electronic states and vibrations. It obeys a symmetry rule:}
\begin{equation}
\label{eq:symmetry1}
\chi^{(3)*}_{K_{pu}}(\omega_{pr};\omega_{pr}',-\omega_{pu}',\omega_{pu}) = \chi^{(3)}_{K_{pu}}(-\omega_{pr};-\omega_{pr}',\omega_{pu}',-\omega_{pu})
\end{equation}
analogously to the symmetry rule for its Cartesian form in eq.~(\ref{eq:symmetry}).

Comparing eqs.~(\ref{eq:A(Kpu)}), (\ref{eq:P(Kpu)}), and (\ref{eq:symmetry1}) we conclude that the coefficients $ A_{K_{pu}}(\omega_{pr})$ in eq.~(\ref{eq:general}) obey the symmetry rule

\begin{equation}
\label{eq:symmetry2}
A^*_{K_{pu}}(\omega_{pr})= -A_{K_{pu}}(-\omega_{pr}).
\end{equation}


An important symmetry rule for nonlinear susceptibility $\chi^{(3)}_{K_{pu}}(\omega_{pr};\omega_{pr}',-\omega_{pu}',\omega_{pu})$ can be obtained by close inspection of  eqs.~(\ref{eq:chi(3)})--(\ref{eq:D}).   If the molecule under study obeys the reflection symmetry, the electron-vibrational wavefunctions $|g\rangle$,   $|e\rangle$,  $|e'\rangle$,  $|f\rangle$ in the transition matrix elements in  eq.~(\ref{eq:D}) can be assumed to belong to the fundamental irreducible representation $^1\!A_1$ and remain the same after reflection of all electron coordinates in the mirror plane whereas the dipole projection $d_{q}$  is transformed to $d_{-q}$.

Therefore, proceeding reflection of all electron coordinates in the dipole integrals $\langle g | d_{q_1} | e \rangle$  and $\langle e | d_{q_2} | f \rangle$   in eq.~(\ref{eq:D}), making  the change of notation $q_1\rightarrow -q_1$,  $q_2\rightarrow -q_2$ in the sum in eq.~(\ref{eq:chi(3)}), and using the symmetry properties of the Chebsch-Gordan coefficient, one can obtain that the expression for the nonlinear  susceptibility  in eq.~(\ref{eq:chi(3)}) is multiplied by a phase factor $(-1)^{K_{pu}}$ and therefore vanishes at $K_{pu}=1$. Therefore in the case of non-chiral molecules the rank $K_{pu}$ is limited by only even values $K_{pu}$\,=\,0 and 2. The former value refers to the number of excited molecules and the latter to the molecular axes alignment by the pump light beam.

The obtained result suggests that the molecular axes can only be aligned under excitation with light, but not oriented. This statement generalises the conclusion made of our recent theoretical study of excitation of symmetric top polyatomic molecules with polarized light and their detection by pump-probe scheme~\cite{Gorbunova20b,Semak21}.  In the following we will limit ourselves to the case of non-chiral molecules and consider only even values of the pump beam rank values $K_{pu}$\,=\,0 and 2.

\subsection*{Components of the Polarization Tensor $E_{KQ}(\mathbf{n})$ of a Light Beam Propagating along Direction $\mathbf{n}$}

Slightly different versions of the expressions  for the components of polarization tensor $E_{KQ}$ in eq.~({\ref{eq:EKQ}}) were tabulated by several authors~\cite{Dyakonov64,Happer72,Zare88b,Alex93,Kupriyanov93b}. However, all of them considered the tensor components of linearly polarized light with respect to the light polarization vector $\mathbf{e}$ and the tensor components of circularly polarized and unpolarized light with respect to the beam propagation direction $\mathbf{n}$. This was not convenient for this paper study of the polarization properties of SRS where both laser beams propagate along axis $Z$.

We consider the Cartesian frame XYZ shown in Fig.~\ref{fig:BDF} and the Beam frame (BF) X$'$Y$'$Z$'$ where $\mathbf{n} \|\mathrm{Z}'$ is the direction of light beam propagation and $\alpha$, $\beta$, $\gamma$ are Euler angles.

General expressions for the tensor components $E_{KQ}$  for arbitrary elliptically polarized light beam propagating along the direction $\mathbf{n}$ can be presented as.

\begin{align}
\label{eq:EPE00}
E_{00}(\mathbf{n})&=-\frac{1}{\sqrt3}  \\
\label{eq:EPE10}
E_{10}(\mathbf{n})&= \frac{1}{\sqrt2}\cos\beta\sin2\gamma\sin\delta  \\
\label{eq:EPE11}
E_{11}(\mathbf{n})&= -E^*_{1-1}(\mathbf{n})  = -\frac{1}{2}e^{i\alpha}\sin\beta\sin2\gamma\sin\delta  \\
\label{eq:EPE20}
E_{20}(\mathbf{n})&= \frac{1}{\sqrt6}\left[\frac{3}{2}\sin^2\beta\cos2\gamma-\frac{1}{2}(3\cos^2\beta-1) \right]  \\
\label{eq:EPE21}
E_{21}(\mathbf{n})&= -E^*_{2-1}(\mathbf{n}) =\,e^{i\alpha}\,\sin\beta\cos\gamma  \nonumber \\
&\times
\Big[\cos\beta\cos\gamma+i\sin\gamma\cos\delta\Big]  \\
E_{22}(\mathbf{n})
&= E^*_{2-2}(\mathbf{n}) = \frac{1}{2}e^{2i\alpha}\Big[\cos^2\beta\cos^2\gamma   \nonumber \\
&
-\sin^2\gamma+i\cos\beta\sin2\gamma\cos\delta\Big].\label{eq:EPE22}
\end{align}

\begin{scheme} [h]
   \centering
   \includegraphics[height=5cm]{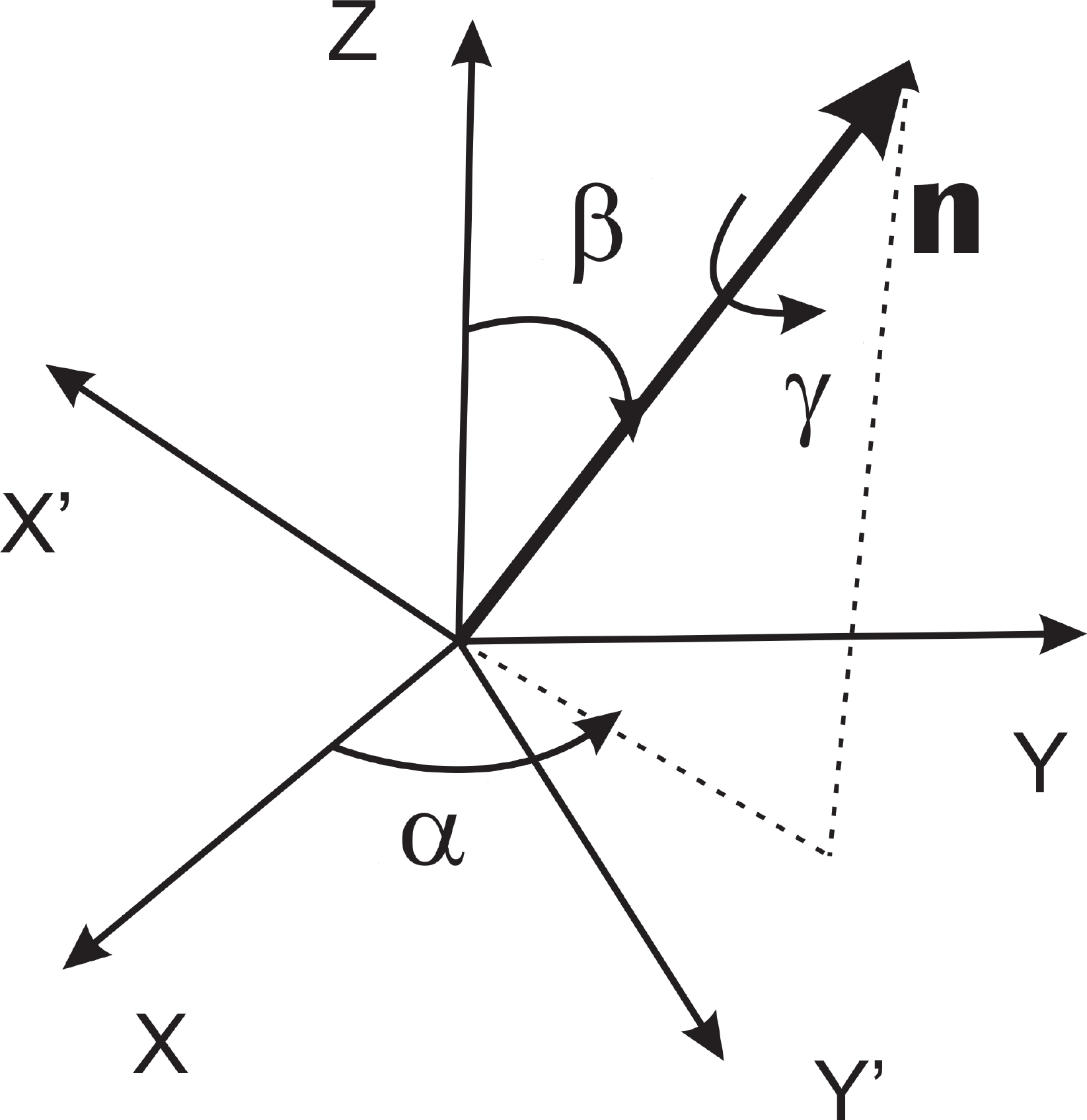}
   \caption{Laser beam polarization frame}
    \label{fig:BDF}
\end{scheme}

In eqs.~(\ref{eq:EPE00})--(\ref{eq:EPE22}) $\alpha$ and $\beta$  are two Euler angles defining the direction of beam propagation $\mathbf{n}$  in the laboratory frame XYZ, $\gamma$  is the angle specifying the direction of the light polarization vector $\mathbf{e}$  in the beam frame X$'$Y$'$Z$'$ where $\mathbf{n}\|\mathrm{Z}'$, and $\delta$ is the phase shift between light polarization components $\mathbf{e}_{X'}$ and $\mathbf{e}_{Y'}$.

In particular in the case of linearly polarized light $\delta=0$ and in case of right circularly polarized light $\delta=\pi/2$ and $\gamma=\pi/4$. In the case of circularly polarized light eqs.~(\ref{eq:EPE00})--(\ref{eq:EPE22}) can be readily reduced to the earlier tabulated expressions~\cite{Dyakonov64,Zare88b,Alex93,Kupriyanov93b}.

If the light beam propagates along the laboratory frame Z, $\beta=0$ and eqs.~(\ref{eq:EPE00})--(\ref{eq:EPE22}) can be simplified as:
\begin{align}
\label{eq:EPE00Z}
E_{00}(\mathbf{n})&=-\frac{1}{\sqrt3}  \\
\label{eq:E10}
E_{10}(\mathbf{n})&= \frac{1}{\sqrt2}\sin2\gamma\sin\delta  \\
E_{11}(\mathbf{n})&= 0  \\
\label{eq:EPE20Z}
E_{20}(\mathbf{n})&= -\frac{1}{\sqrt6}  \\
E_{21}(\mathbf{n})&= 0  \\
E_{22}(\mathbf{n})&= E^*_{2-2}(\mathbf{n}) = \frac{1}{2}\Big[\cos2\gamma+i\sin2\gamma\cos\delta\Big].\label{eq:EPE22Z}
\end{align}

As can be seen in eqs.~(\ref{eq:EPE00Z})--(\ref{eq:EPE22Z}) the only meaningful polarization tensor components that can be changed after the light beam passed through the molecular sample are: $E_{10}(\mathbf{n})$ and $E_{2\pm 2}(\mathbf{n})$. Using eqs.~(\ref{eq:Stokes2}), (\ref{eq:E10}), and (\ref{eq:EPE22Z}) one can readily express the Stokes parameters $S_1$, $S_2$, $S_3$ in terms of the angle $\gamma$ and the phase shift $\delta$ as follows:

\begin{align}
\label{eq:StokesS1}
S_1&=\frac{I_x-I_y}{I_x+I_y}=   \cos2\gamma \\
\label{eq:StokesS2}
S_2&=\frac{I_{\pi/4}-I_{(-\pi/4)}}{I_{(\pi/4)}+I_{(-\pi/4)}}= \sin2\gamma\cos\delta \\
\label{eq:StokesS3}
S_3&=\frac{I_{+}-I_{-}}{I_{+}+I_{-}}= \sin2\gamma\sin\delta,
\end{align}
where $I_x$ and $I_y$ are intensities of light beam polarized along $X$ and $Y$ axes, respectively,  $I_{\pi/4}, I_{-\pi/4}$ are intensities of light beams polarized on $\pm \pi/4$ to axis X, and $I_+$ and $I_-$ are intensities of right and left circularly polarized light beams.

\subsection*{Analysis of the Probe Beam Polarization Tensor Components in Terms of the Stokes Parameters}

\subsubsection*{Probe Light Beam Intensity Gain/Loss}
As is known~\cite{Boyd20} in the presence of a more intensive pump beam the probe Stokes light beam undergoes a gain after passing through the sample due to nonlinear energy transformation. For analysis of the probe beam intensity change after passing through the MS the polarization rank $K$ and its projection $Q$  in eq.~(\ref{eq:general}) should be set to zero. According to the symmetry property of the 6-$j$ symbol in eq.~(\ref{eq:general}) in this case $K_{pu}=K_{pr}$ and the phase $(-1)^{\phi}$ is equal to $+1$.

Substituting $E_{00}(\mathbf{n}) =-1/\sqrt3$ from eq.~(\ref{eq:EPE00Z}) to eq.~(\ref{eq:general}) the beam intensity change
$\Delta I(\omega_{pr})= I^{out}(\omega_{pr}) - I^{in}(\omega_{pr})$ for the case when the pump and probe beams are both linearly polarized can be presented in the form:

\begin{multline}
\label{eq:gain}
    \Delta I(\omega_{pr})= \omega^{(c)}_{pr}\ell\, \mathrm{Im}\Big({\cal E}^*(\omega_{pr}) \\
     \times \Big(P^{(3)}_{0}(\omega_{pr})+\frac{1}{2}P^{(3)}_{2}(\omega_{pr})+\frac{3}{2}\cos(2\Delta^{(in)}\gamma)
                  P^{(3)}_{2}(\omega_{pr})\Big)\Big),
\end{multline}
where $\Delta^{(in)}\gamma = \gamma_{pu}-\gamma_{pr}$ is the angle between the polarization directions of the incoming pump and probe beams.

The probe beam intensity change $ \Delta I(\omega_{pr})$ in eq.~(\ref{eq:gain}) is presented in a traditional form $\Delta I \sim \mathrm{Im}[{\cal E}^* P^{(3)}]$~\cite{Mukamel95,Boyd20}, however the polarization components are specified in more details.  As can be seen in eq.~(\ref{eq:gain}) the nonlinear 3-rd order polarization contains contributions from three terms. The term $P^{(3)}_{0}$ represents contribution from isotropic absorption/gain of the probe beam, while two other terms represent different types of contributions from molecular axes alignment.  The second term in eq.~(\ref{eq:gain}) refers to the longitudinal alignment of the molecular axes that is parallel to the laser beam direction. The third term in eq.~(\ref{eq:gain}) refers to the transversal alignment of the molecular axes that is perpendicular to the laser beam direction. As shown in eq.~(\ref{eq:gain}) the latter term depends on the relative initial polarizations of the pump and probe laser beams and vanishes at the Kerr geometry ($\Delta^{(in)}\gamma) = \pi/4$).

\subsubsection*{Rotation of the Probe Beam Polarization}

We consider the case when both pump and probe beams are linearly polarized and a linear analyser $A$ is installed in front of the photodetector, see Scheme~\ref{fig1}. As shown in eqs.~(\ref{eq:EPE00Z}), (\ref{eq:EPE20Z}), and (\ref{eq:EPE22Z}) the nonzero components of the polarization tensor for each beam are as follows: $E_{00}=-1/\sqrt3$, $E_{20}=-1/\sqrt6$, $E_{2\pm2}=1/2\,e^{\pm 2i\gamma}$, where the polarization angle $\gamma$ is either $\gamma_{pu}$, or $\gamma_{pr}$. According to eq.~(\ref{eq:EPE22Z}) the rotation of the probe beam polarization plane is described by the tensor component $E_{2\pm2}$. Substituting $K=2$, $Q=2$ into eq.~(\ref{eq:general}) and transforming it having in mind eq.~(\ref{eq:Stokes2}) the expressions describing the Stokes parameter $S^{out}_1$ of the outgoing probe beam can be written in the form:

\begin{multline}
\label{eq:E22-3}
I^{out}(\omega_{pr})S^{out}_1
         = I^{in}(\omega_{pr})S^{in}_1
         -\frac{2}{3}\Big(
             \,S^{in}_1 \mathrm{Re}[A^{}_{0}(\omega_{pr})] \\
             + S^{in}_1\frac{1}{2} \mathrm{Re}[A^{}_{2}(\omega_{pr})]
        +
           \frac{3}{2}\,S^{pu}_1  \mathrm{Re}[A^{}_{2}(\omega_{pr})]
         \Big)
         \end{multline}
where $S^{pu}_1$ and $S^{in}_1$ and the Stokes parameters of the incoming pump and probe beams, respectively.

Equation (\ref{eq:E22-3}) is valid for arbitrary linear polarization of each of the probe and pump beams propagating collinearly along axis $Z$. In the case of Kerr geometry when the pump beam is linearly polarized along axis $X$ and the probe beam is linearly polarized on 45$^{\circ}$ to axis X, $S^{pu}_1$=1 and $S^{in}_1$=0. Substituting these values into eq.~(\ref{eq:E22-3}) it can be transformed to a simple and physically clear form:
          \begin{equation}
\label{eq:E22-3K}
S^{out}_1 =  -\frac{\mathrm{Re}[{A}^{}_{2}(\omega_{pr})]}{I^{out}(\omega_{pr})}=\frac{\omega^{(c)}_{pr}\ell\,\mathrm{Im}\left[{\cal E}^*(\omega_{pr}) P_{2}(\omega_{pr})\right]}{I^{out}(\omega_{pr})}.
          \end{equation}

According to eq.~(\ref{eq:StokesS1}) the Stokes parameter $S_1$ is equal to the
\begin{equation}
\label{eq:S1}
S^{out}_1 = \cos2\gamma = \sin(2\Delta\gamma),
          \end{equation}
          where $\Delta\gamma=(\pi/4-\gamma)$ is the angle of rotation of the probe beam polarization after passing through MS.

The physical reason for rotation of the light polarization vector is the linear dichroism of MS for the probe beam passing through it~\cite{Semak21}.

\subsubsection*{Probe Beam Birefringence}

As shown in eqs.~(\ref{eq:StokesS2}) and (\ref{eq:StokesS3}) the Stokes parameters $S_2$ and $S_3$ contain the phase shift $\delta$ between the orthogonal light polarization components $E_x$ and $E_y$. This phase shift can be created in the  probe beam passing through anisotropic media. We again consider the case when both pump and probe beams are linearly polarized, but now assume a circular analyser $A$ installed after the MS in front of the D, see Scheme~\ref{fig1}. According to eq.~(\ref{eq:E10}) the phase shift $\delta$ is contained in the matrix element $E_{1 0}$ related to the Stokes parameter $S_3$ in eq.~(\ref{eq:Stokes2}). Substituting $K=1$, $Q=0$ into eq.~(\ref{eq:general}) and transforming it having in mind eq.~(\ref{eq:Stokes2}) the expressions describing the Stokes parameter $S^{out}_3$ of the outgoing probe beam can be written in the form:
\begin{multline}
\label{eq:S3-1}
 I^{out}(\omega_{pr})S^{out}_3                 =
 \mathrm{Im}(A^{}_{2}(\omega_{pr})) \,
                  \Big(\sin2\gamma_{pu}\cos\delta_{pu}\cos2\gamma^{in}_{pr}\\
                  -\sin2\gamma^{in}_{pr}\cos\delta^{in}_{pr}\cos2\gamma_{pu}\Big).
        \end{multline}

 Having in mind that for incoming linearly polarized beams $\delta_{pu}=\delta^0_{pr}=0$ and using Kerr geometry: $\gamma_{pu}=0$, $\gamma^{in}_{pr}=\pi/4$ eq.~(\ref{eq:S3-1}) is simplified to the form:

\begin{equation}
             \label{eq:CPprobe1}
S^{out}_3=  -\frac{\omega^{(c)}_{pr}\ell\,\mathrm{Re}\left[{\cal E}^*(\omega_{pr}) P_{2}(\omega_{pr})\right]}{I^{out}(\omega_{pr})}.
           \end{equation}

Therefore, determination of the Stokes parameters $S_1$~(\ref{eq:StokesS1}) and $S_3$~(\ref{eq:StokesS3}) in the Kerr geometry allows to determine both the polarization rotation angle $\Delta\gamma$ and the phase shift $\delta$ from experiment.

\subsection*{Particular Cases of Third-Order Nonlinear Susceptibility}

The nonlinear susceptibility $\chi^{(3)}_{K_{pu}}(\omega_{pr};\omega_{pr}',-\omega_{pu}',\omega_{pu})$
in   eq. (\ref{eq:chi(3)}) and the coefficients $A_{K_{pu}}(\omega_{pr})$ in eq.~(\ref{eq:A(Kpu)}) are in general complex functions of frequencies.

The function $R_{\tilde q}(\tilde\omega)$ in eqs.~(\ref{eq:chi(3)}) and (\ref{eq:R}) contains the details of molecular energy levels, different excitation schemes, relaxation and dephasing rates, and optical transitions occurring. In general this function contains 48 terms.  However, as known~\cite{Boyd20}  only a few of them can give significant contributions to the signal, these are the terms containing resonances. At the same time as can be seen in the first and the second lines eq.~(\ref{eq:general}) all  possible excitation schemes are described by the universal light polarization dependence that is an undeniable advantage of the spherical tensor approach. Here we consider two multiphoton excitation schemes that are relevant to the conditions of our recent experiments~\cite{Gorbunova20b}.

One of these schemes is off-resonance SRS qualitatively shown in Scheme~\ref{fig:Scheme 2} where the outgoing probe beam is denoted by the bold downward arrow $\mathbf{pr}$.

\begin{scheme} [h]
   \centering
   \includegraphics[width=6.0cm]{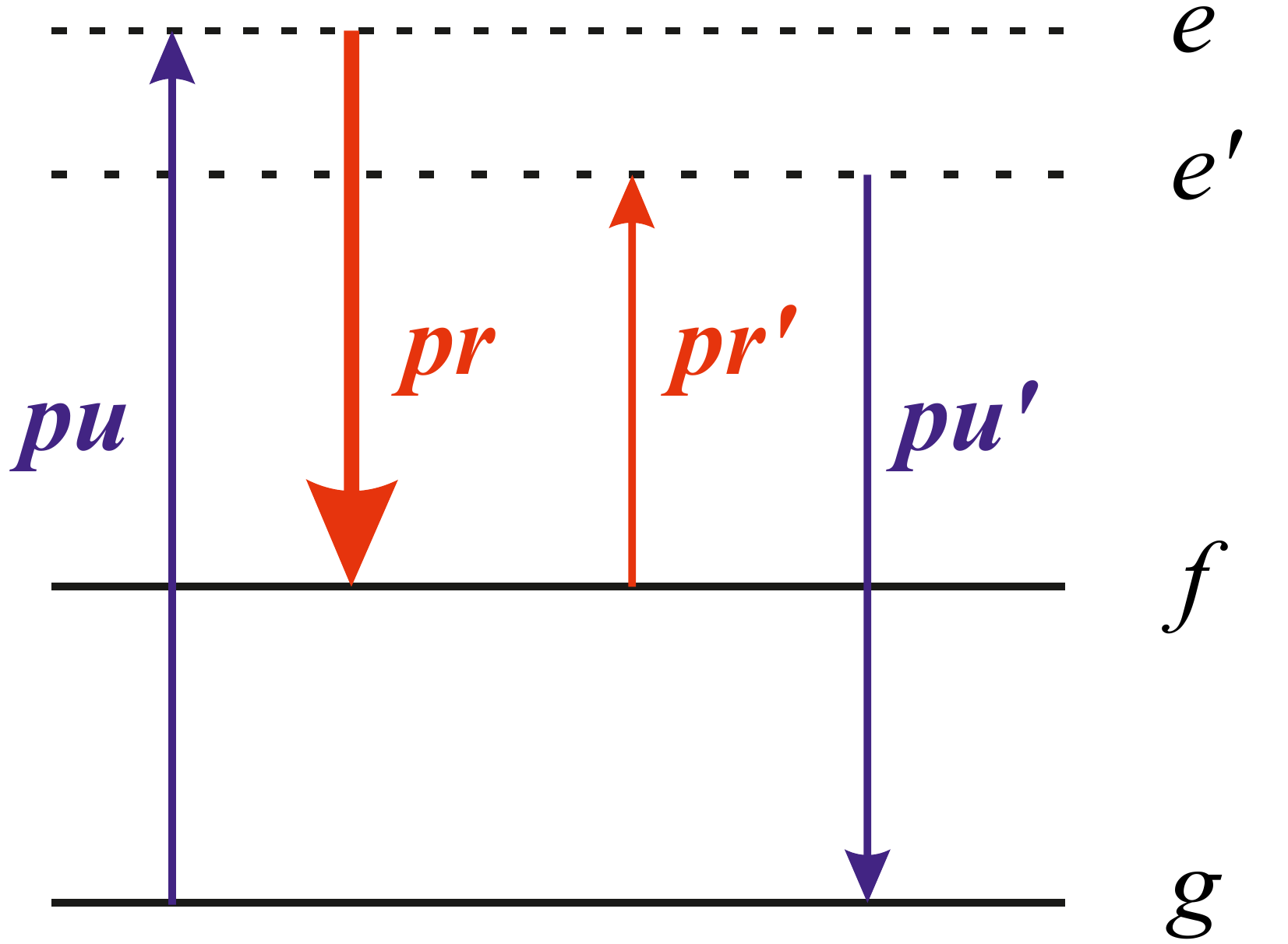}
   \caption{The scheme of off-resonance SRS.}
    \label{fig:Scheme 2}
\end{scheme}

Here $|g\rangle$ and $|f\rangle$ are molecular ground and final states, and $|e\rangle, |e'\rangle$ are intermediate states.  It is assumed that the energies of the excited electronic states $|e\rangle$ and $|e'\rangle$ are much larger than the laser photon energies: $\omega_{eg},\omega_{e'g}\gg \omega'_{pu},\omega_{pu},\omega'_{pr},\omega_{pr}$, while the state $|f\rangle$  belongs to the same electronic ground state as $|g\rangle$.   Moreover, usually in experiment the probe beam intensity is detected as a whole without any frequency resolution. Then, the general expression for the change of the probe beam polarization matrix in eq.~(\ref{eq:general}) should be integrated over the frequency $\omega_{pr}$.  All these assumptions perfectly met the conditions of many SRS experiments utilising laser impulses with duration of tens, or hundreds of femtoseconds.

In Scheme~\ref{fig:Scheme 2} the frequency difference $(\omega_{pu'}-\omega_{pr'})$ is assumed to be positive and much larger than the beams frequency envelope widths. In this case the major contribution to the SRS signal is given by the terms  $I^*_{eg}(-\omega_{pr}) \ I^*_{fg}(\omega_{pu}'-\omega_{pr}') I^*_{e'g}(-\omega_{pr'})$  and $I_{e'f}(\omega_{pr}) \  I_{gf}(\omega_{pr}'-\omega_{pu}') I_{ge}(-\omega_{pu}')$  in  eq.~(\ref{eq:I_gee'f}). Each of them contains a resonance at $\omega_{fg} \approx (\omega_{pu'}-\omega_{pr'})$  and two off-resonant factors that can only slightly depend on the laser beam frequencies. The contribution from the  former term to the nonlinear susceptibility   in eq.~(\ref{eq:chi(3)}) can be estimated as:

\begin{multline}
\label{eq:chi(3)-1}
\chi^{(3)}_{K_{pu}} \propto \, \sum_{gee'f}\sum_{\tilde q}  (-1)^{q}\,  C_{1\, q_3\,1\,q_4}^{K_{pu}q}\,C_{1\,-q_2\,1\,-q_1}^{K_{pu}q}
\\
\times
\frac{\rho_{gg} \,D_{\tilde q}(g,e,e',f)}{(\omega^{(c)}_{pr}+\omega_{eg})(\omega^{(c)}_{pr}+\omega_{e'g})}
 \frac{1}{[(\omega'_{pu}-\omega'_{pr})-\omega_{fg}-i\Gamma_{fg}]},
\end{multline}
where we neglected the dephasing factors $i\Gamma$  in the off-resonant terms and replaced there the frequencies $\omega_{pr}$ and $\omega'_{pr}$  by the carrier frequency $\omega^{(c)}_{pr}$.

Assuming that the dephasing rate $\Gamma_{fg}$ in eq.~(\ref{eq:chi(3)-1}) is much smaller than the frequency difference  $\Gamma_{fg} \ll (\omega'_{pu}-\omega'_{pr})$  one should expect a sharp resonance at the energy $\omega_{fg}\approx (\omega'_{pu}-\omega'_{pr})$  in the molecular vibrational energy quasi-continuum.  Integrating eq.~(\ref{eq:chi(3)-1}) over the quasi-continuum frequency $\omega_{fg}$ from $-\infty$  to $\infty$  the nonlinear susceptibility $\chi^{(3)}_{K_{pu}}$ can be shown to have a pure imaginary value:

\begin{multline}
\label{eq:chi(3)-2}
\chi^{(3)}_{K_{pu}} \propto i\pi\sum_{gee'f}\sum_{\tilde q}  (-1)^{q}\,  C_{1\, q_3\,1\,q_4}^{K_{pu}q}\,C_{1\,-q_2\,1\,-q_1}^{K_{pu}q}
\\
\times
\frac{\rho_{gg} \,D_{\tilde q}(g,e,e',f)}{(\omega^{(c)}_{pr}+\omega_{eg})(\omega^{(c)}_{pr}+\omega_{e'g})}.
\end{multline}

The vibrational wavefunctions $\Psi_{v_f}$ in $D_{\tilde q}(g,e,e',f)$ in eq.~(\ref{eq:chi(3)-2}) should be calculated at the energy $\omega_f=\omega_g+|\omega^0_{pu}-\omega^0_{pr}|$.

Substitution of eq.~(\ref{eq:chi(3)-2}) into eqs.~(\ref{eq:P(Kpu)}) and (\ref{eq:gain}) and proceeding the Fourier transform over the frequency $\omega_{pr}$  and integration  over the electric field frequencies $\omega_{pu}$, $\omega'_{pu}$, and $\omega'_{pr}$  results in the time-dependence of the probe beam intensity gain  in eq.~(\ref{eq:gain}) in the form:
\begin{equation}
\label{eq:int}
\langle \Delta I(\omega^{(c)}_{pr})  \rangle
\propto
\omega^{(c)}_{pr}\ell \,\sigma^4\,e^{-\frac{\sigma^2\tau^2}{2}},
\end{equation}
where $\tau$ is the delay time between the pump and probe pulses, which envelopes are assumed to be Gaussian with the width $\sigma$.The angular brackets denote averaging over fast oscillations with the frequency  $\omega^{(c)}_{pr}$.

According to eqs.~(\ref{eq:gain}), (\ref{eq:E22-3}), and (\ref{eq:CPprobe1}) in the scheme of off-resonance SRS shown in Scheme~\ref{fig:Scheme 2} the coefficients $\langle A^{}_{K_{pu}}(\omega) \rangle $  can result in the gain of the probe beam intensity in eq.~(\ref{eq:int}) and in the rotation of the probe light polarization vector described by the Stokes parameter $S_1$ in eq.~(\ref{eq:E22-3K}). However the birefringence of the probe beam polarization, see eq.~(\ref{eq:CPprobe1}) is damped in this case.

\begin{scheme} [h]
   \centering
   \includegraphics[width=6.0cm]{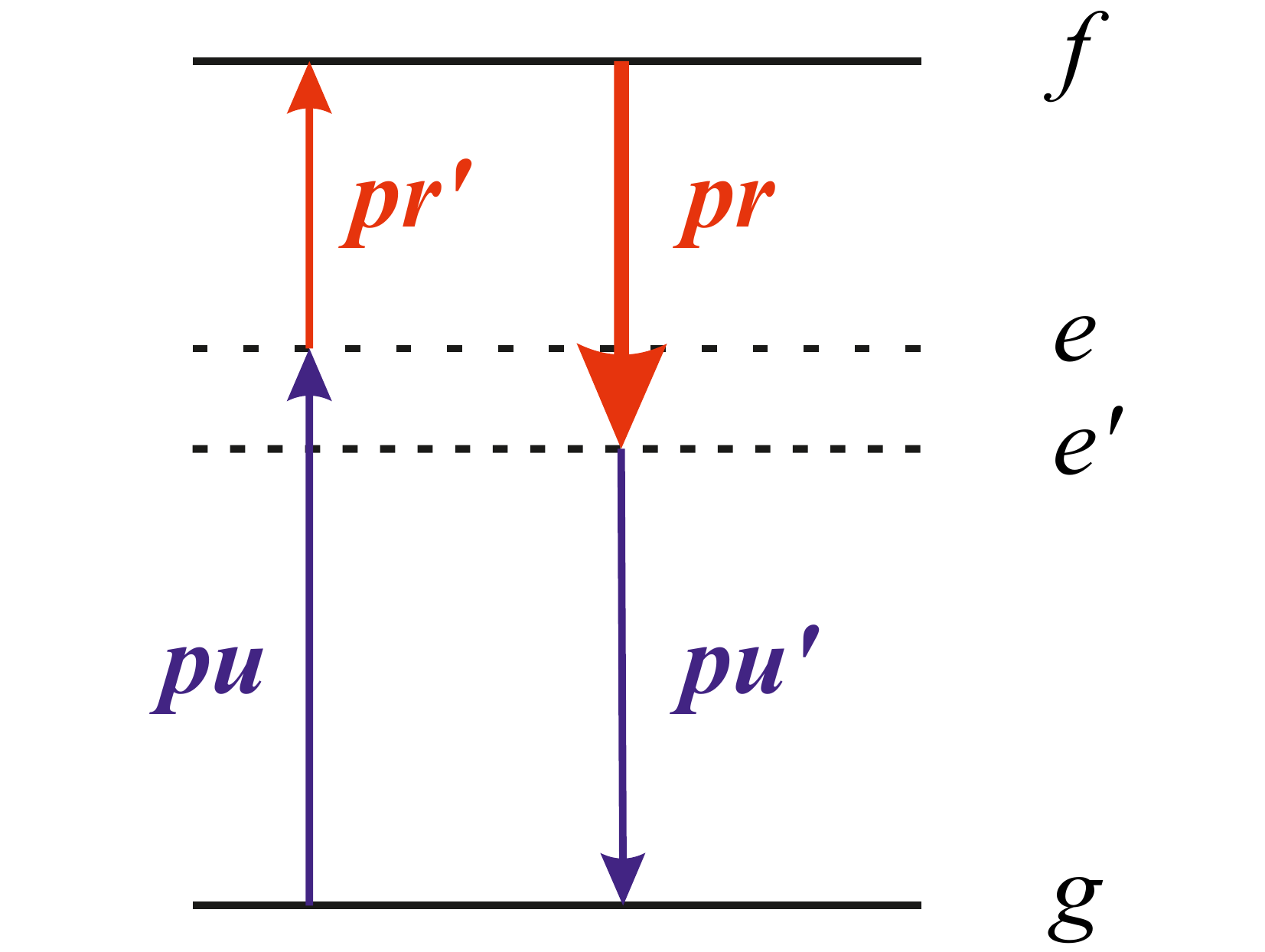}
   \caption{The scheme of resonance pump-probe}
\label{fig:Scheme 3}
\end{scheme}

Another scheme is resonance pump-probe (PP) that is shown in Scheme~\ref{fig:Scheme 3}  where  the outgoing beam is denoted by the bold downward arrow $\mathbf{pr}$. In this case the pump beam is in resonance with the states $|e\rangle$ and $|e'\rangle$ while the probe beam can be assumed either to be in resonance or off-resonance with the state $|f\rangle$.

In this case the major contribution to the PP signal is given by the terms  $I_{e'f}(\omega_{pr}) I_{e'e}(\omega_{pu}-\omega_{pu}') I_{e'g}(\omega_{pu})$ and $I_{e'f}(\omega_{pr}) I_{e'e}(\omega_{pu}-\omega_{pu}') I_{ge}(-\omega_{pu}')$   in  eq.~(\ref{eq:I_gee'f}) and their complex conjugated counterparts.  If the inverse lifetimes $\gamma_{e}$ and $\gamma_{e'}$ are larger than the  pure dephasing rate $\hat{\Gamma}_{e,e'}$ (see eq.~(\ref{eq:relax})) the contribution from these terms to the nonlinear susceptibility   in eq.~(\ref{eq:chi(3)}) can be estimated as:

\begin{multline}
\label{eq:chi(3)-4}
\chi^{(3)}_{K_{pu}} \propto  \sum_{gee'f}\sum_{\tilde q}  (-1)^{q}\,  C_{1\, q_3\,1\,q_4}^{K_{pu}q}\,C_{1\,-q_2\,1\,-q_1}^{K_{pu}q}
  D_{\tilde q}(g,e,e',f)\\
\times \Bigg(
\frac{1}{(\omega_{pr}-\omega_{fe}+i\Gamma_{ef})}\\
\times
\frac{1}{(\omega_{pu}+\omega_{eg}+i\Gamma_{eg})
(\omega'_{pu}+\omega_{e'g}-i\Gamma_{e'g})} \\
-
\frac{1}{(\omega_{pr}+\omega_{fe'}+i\Gamma_{e'f})}
\\
\times
 \frac{1}{(\omega_{pu}-\omega_{e'g}+i\Gamma_{e'g})
(\omega'_{pu}-\omega_{eg}-i\Gamma_{eg})} \Bigg).
\end{multline}

In general, the nonlinear susceptibility in eq.~(\ref{eq:chi(3)-4}) can have real and imaginary parts that relate to the real and imaginary values of the third-order nonlinear polarization $P^{(3)}_k(\omega_{pr})$ in eq.~(\ref{eq:P(Kpu)}) and of the coefficients $A^{}_{K_{pu}}(\omega)$ in eq.~(\ref{eq:general}). The imaginary part of the nonlinear susceptibility refers to the gain of the probe beam in eq.~(\ref{eq:gain}) and to the rotation of the probe beam polarization described by the Stokes parameter $S_1$ in eqs.~(\ref{eq:E22-3}) and (\ref{eq:E22-3K}). According to the results of our recent publication~\cite{Semak21} the latter effect relates to the linear dichroism of the probe beam.

The real part of the nonlinear susceptibility in eq.~(\ref{eq:chi(3)-4}) refers to the probe beam birefringence described by the Stokes parameter $S_3$ in eqs.~(\ref{eq:S3-1}) and (\ref{eq:CPprobe1}), see our recent publication~\cite{Semak21} for details.

\section*{Conclusion}
\label{conclusion}

Full quantum mechanical expressions have been derived describing the change of the polarization tensor of the probe light  after transmission through a molecular sample for any initial polarization states of the pump and probe laser beams. The pump and probe beams were assumed to be pulsed with pulse frequency cover widths much smaller than the beam carrier frequencies. Spherical tensor operators were used throughout the paper that allowed to consider arbitrary angular molecular momenta and to present  the general expression for the change of the probe beam polarization tensor as a product of a polarization-dependent field tensor and a material (molecular) part containing scalar values of the third-order nonlinear optical susceptibility $\chi^{(3)}_{K_{pu}}$ with the pump light rank $K_{pu}$  limited to the values $0,1,2$. The expressions are valid for arbitrary directions of propagation of both pump and probe light beams and their arbitrary polarizations. It was shown that  only two rank values $K_{pu} =0,2$ can contribute to the SRS signal from non-chiral polyatomic molecules. The geometry of almost collinear propagation of the pump and probe pulse beams through the molecular sample was considered in greater details. The expressions contain contributions from the probe beam intensity gain/loss,   linear dichroism, and birefringence that can be described in terms of the Stokes polarization parameters. It was shown that both dichroism and birefringence refer to the nonlinear optical susceptibility element $\chi^{(3)}_{2}$  and that their contributions to the signal  can be separated experimentally by using an appropriate probe beam polarization analyzer installed in front of the photodetector.  In the case of the off-resonance SRS when  the energies of the excited electronic states are much larger than the laser photon energies the nonlinear optical susceptibility has almost pure imaginary value resulting in the gain of the probe beam intensity and in the rotation of the probe light polarization due to linear dichroism described by the Stokes parameter $S_1$. In the case of the resonance SRS when  the pump beam frequency is in resonance with the electronic transitions the nonlinear optical susceptibility  in general has real and imaginary parts while the former  refers to the probe beam birefringence described by the Stokes parameter $S_3$.



\section*{Acknowledgements}

YMB is grateful to the Ministry of Science and Higher Education of the Russian Federation for financial support within the World-class Research Center program Advanced Digital Technologies, Contract No.~075-15-2022-311 from 20.04.2022.

\section*{Conflict of Interest}

There is no any conflict of interest to declare.

\begin{shaded}
\noindent\textsf{\textbf{Keywords:} \keywords{Stimulated Raman Scattering, polarization matrix, spherical tensor representation, dichroism, birefringent} \    }
\end{shaded}



\setlength{\bibsep}{1.0cm}
\bibliographystyle{Wiley-chemistry}
\bibliography{SRS}

\clearpage

\noindent\rule{11cm}{2pt}
\begin{minipage}{11cm}
\includegraphics[width=11cm]{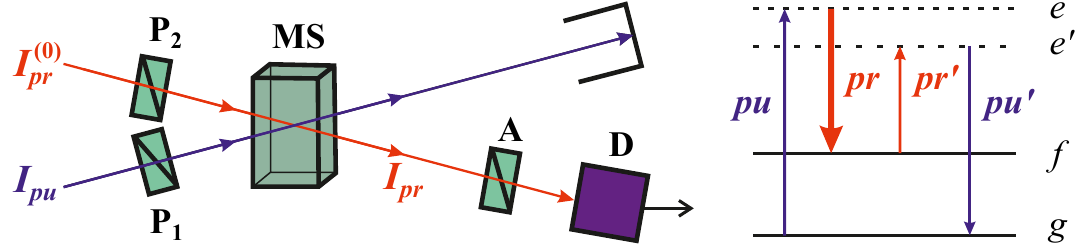}
\end{minipage}
\begin{minipage}{11cm}
\large\textsf{The contributions of the dichroism and birefringence to the SRS signal depend strongly on the energy level structure of the molecular sample. They can be separated experimentally by using an appropriate probe beam polarization analyzer installed in front of the photodetector (D).}
\end{minipage}
\noindent\rule{11cm}{2pt}

\end{document}